# Exploring Multivariate Data Using Median Absolute Deviation Depth

**Elsayed A. H. Elamir**
Department of Management and Marketing, College of Business Administration, Kingdom of Bahrain
Email: shabib@uob.edu.bh

**Abstract**

This paper studies the moving median absolute deviation as a functional description of half mass concentration in univariate and multivariate distributions. Its minimizing location defines the median radius centre. Because the original radius profile need not increase outward, a radial running maximum envelope is constructed. Under continuity and unique minimizer assumptions, normalization yields a continuous centrality score with maximality, outward monotonicity and vanishing values at infinity. Euclidean distance provides similarity invariance, and a univariate derivative identity characterizes local boundary imbalance. Empirical constructions and numerical illustrations distinguish concentration geometry from centre-outward ordering and clarify the interpretation of central regions under asymmetric and multimodal distributions.

ORCID: 0000-0002-9430-072X

# 1 Introduction

Robust scale measures play an important role in univariate and multivariate analysis by providing summaries of dispersion that are less sensitive to extreme observations. The median absolute deviation is a prominent example with established breakdown and influence function properties under appropriate conditions (Maronna et al., 2018; Rousseeuw & Croux, 1993). Multivariate extensions similarly combine a measure of distance with a selected location. Such measures are commonly reported as global summaries evaluated at a fixed centre rather than examined as functions of a varying candidate location (Arachchige & Prendergast, 2026; Capezza et al., 2024).

Centre-outward measures provide a complementary representation by ordering locations according to their centrality relative to a reference distribution. Statistical depth offers an established framework for such orderings with desirable properties including invariance, maximality at a centre, monotonicity away from a deepest point and vanishing at infinity. when the distribution contains asymmetric or separated concentrations, a concentration descriptor does not automatically satisfy these requirements (Rousseeuw & Leroy, 1987; Serfling & Zuo, 2000).

This paper considers the moving median absolute deviation. For a candidate location $v$, the univariate functional is the median of $|X - v|$ while its multivariate counterpart is the median of $\|X - v\|$. Each value represents the smallest radius of a neighbourhood centred at $v$ containing at least half the probability mass. Evaluated at a distributional median in one dimension, the functional reduces to the unscaled median absolute deviation (Rousseeuw & Leroy, 1987; Schwank & Drton, 2026). Allowing the candidate location to vary produces a profile of half mass concentration rather than a single scale summary. The minimizing location is termed the median radius centre (MRC). In one dimension, it is the midpoint of a shortest interval containing half the probability mass. This criterion is connected to established shortest half and least median methods and is not introduced here as a new location principle (Capezza et al., 2024; E., 2025; Rousseeuw & Leroy, 1987). The present development examines the complete moving-radius profile, its centre-outward modification, and its boundary behaviour. Recent work on the geometric median examines both statistical estimation and numerical approximation including applications to robust mean estimation. These results concern a different location criterion from the distance quantile criterion studied here (Minsker & Strawn, 2024).

Under the stated assumptions, the original median radius function is continuous but need not be convex or monotone away from its minimizer. Consequently, its normalized reciprocal does not automatically provide a centre-outward ordering. We address this limitation by taking the running maximum of the radius along each segment joining the selected MRC to a candidate location. The resulting radial monotone envelope is the smallest majorant of the original radius that is nondecreasing along every outward ray. Dividing the minimum radius by this envelope produces a normalized centrality score.

The theoretical results are established on a distribution class with a unique population MRC. On this class, the resulting score is continuous, strictly maximized at the MRC, nonincreasing along outward rays and vanishes at infinity. In multivariate dimensions, Euclidean distance yields similarity invariance and the corresponding multivariate central regions are compact, nested and star shaped although convexity is not guaranteed.

The paper makes two contributions. First, it establishes the basic properties of the moving median radius function and its radial monotone envelope under explicit assumptions governing centre selection and invariance.

Second, it develops an envelope-based MMAD depth measure that provides a continuous centre-outward ordering with strict maximality at the unique population MRC, outward monotonicity and vanishing values at infinity. It characterizes the associated central regions, provides empirical constructions and numerical illustrations.

The remainder of the paper is organized as follows. Section 2 develops the univariate median radius function, its envelope, the associated depth and the boundary interpretation. Section 3 extends the construction to multivariate data and empirical implementation. Section 4 concludes with limitations and directions for further investigation.

# 2 Moving median absolute deviation (MMAD)

Let $X$ be a real valued random variable with continuous distribution function $F$. For any candidate location $v \in \mathbb{R}$, the moving median absolute deviation (MMAD) is defined by

$$G_F(v) = \text{Med}|X - v| = \inf\{r \geq 0: P(|X - v| \leq r) \geq 0.5\}.$$

The median is interpreted as the lower 0.5-quantile. Thus, $G_F(v)$ is the smallest radius of an interval centred at $v$ containing at least half the probability mass. Continuity of $F$ implies

$$F\big(v + G(v)\big) - F\big(v - G(v)\big) = 0.5,$$

When $v$ is a distributional median, $G_F(v)$ equals the unscaled median absolute deviation about that median. This distinguishes the present radius convention from versions of the MAD multiplied by a normal consistency factor (Elamir, 2025; Rousseeuw & Croux, 1993). Allowing the reference location to vary produces a profile of half mass concentration across the real line. Its minimum is connected to established shortest half and least median constructions (Grubel, 1988; Lange et al., 2014). The present development separates two objectives: describing concentration geometry and constructing a centre-outward ordering. These objectives require separate treatment because the original radius need not increase monotonically away from its minimizing centre.

## 2.1 Basic properties and the median radius centre

The following properties establish the existence of a minimizing location without imposing moment conditions.

**Proposition 1.** For a continuous distribution $F$, the functional $G_F$ has the following properties.

1. Positivity: $G(v) > 0$, for every $v \in \mathbb{R}$.
2. Location-scale equivariance: for $Y = a + bX$, where $b \neq 0$,
$$G_{F_Y}(a + bv) = |b| G_F(v).$$
3. Lipschitz continuity: for all $v, w \in \mathbb{R}$, $|G_F(v) - G_F(w)| \leq |v - w|$.
4. Coercivity:
$$G_F(v) \to \infty \qquad \text{as } |v| \to \infty.$$

Consequently, $G_F$ attains a strictly positive minimum and its set of minimizing locations is nonempty and compact. Positivity follows because a continuous distribution assigns zero probability to any singleton. The equivariance relation follows directly from the transformation of absolute distances. The triangle inequality and the ordering of quantiles imply the Lipschitz bound. Coercivity follows because an interval of fixed radius contains probability tending to zero as its centre tends to either infinity.

**Assumption 1.** The population median radius function has a unique global minimizer,

$$\theta_F = \arg\min_{v\in\mathbb{R}} G_F(v),$$

and

$$g_F = G_F(\theta_F) = \min_{v\in\mathbb{R}} G_F(v) > 0$$

We call $\theta_F$ the median radius centre (MRC). The interval

$$[\theta_F - g_F,\ \theta_F + g_F],$$

is a shortest interval containing half the probability mass. Conversely, the midpoint of any shortest half probability interval minimizes $G_F$. This interpretation connects the MRC with shortest half methods and least median of squares location estimation under the corresponding median convention (Rousseeuw & Leroy, 1987; Schwank & Drton, 2026).
while the median describes probability balance, the MRC describes half mass concentration. The two locations therefore need not coincide. Uniqueness is an additional assumption rather than a consequence of continuity.

## 2.2 The center-outward monotone envelope

Under Assumption 1, define the centre-outward monotone envelope by

$$L_F(v) = \max_{t\in[0,1]} G_F\big(\theta_F + t(v - \theta_F)\big).$$

The maximum exists because $G_F$ is continuous and the defining segment is compact.
Equivalently,

$$L_F(v) = \begin{cases} \max\limits_{w\in[v,\theta_F]} G_F(w), & v < \theta_F, \\ G_F(\theta_F), & v = \theta_F, \\ \max\limits_{w\in[\theta_F,v]} G_F(w), & v > \theta_F. \end{cases}$$

Thus, the envelope records the largest radius encountered while moving from the MRC to the candidate location.

**Proposition 2.** Under Assumption 1, $L_F$ satisfies

$$L_F(v) \geq G_F(v),$$

and

$$L_F(\theta_F) = g_F, \qquad L_F(v) > g_F \quad \text{for } v \neq \theta_F.$$

For $u \in \{-1, 1\}$,

$$L_F(\theta_F + r_1 u) \leq L_F(\theta_F + r_2 u), \quad 0 \leq r_1 \leq r_2.$$

Furthermore,

$$|L_F(v) - L_F(w)| \leq |v - w|,$$

and

$$L_F(v) \to \infty \qquad \text{as } |v| \to \infty.$$

The function $L_F$ is the smallest majorant of $G_F$ that is nondecreasing along both outward rays from $\theta_F$. To see this, let $M$ be any such majorant. For every point $w$ between $\theta_F$ and $v$, $G_F(w) \leq M(w) \leq M(v)$. Taking the maximum over the segment gives $L_F(v) \leq M(v)$. The Lipschitz property follows by comparing the defining expressions at the same $t$ and then taking maxima. If $G_F$ is already nondecreasing along both outward rays, we have

$$L_F(v) = G_F(v) \qquad \text{for all } v \in \mathbb{R}.$$

The envelope therefore changes the original profile only where outward monotonicity fails. Its sublevel sets are intervals, hence $L_F$ is quasiconvex in one dimension.

## 2.3 Envelope-based MMAD depth

Define the envelope-based MMAD depth by

$$D_F(v) = \frac{g_F}{L_F(v)}.$$

This dimensionless ratio compares the minimum half mass radius with the envelope radius at the candidate location.

**Theorem 1.** On the class of continuous distributions satisfying Assumption 1, $D_F$ is continuous and satisfies the following properties.

1. Affine invariance in one dimension: for $Y = a + bX$, where $b \neq 0$,
$$D_{F_Y}(a + bv) = D_F(v).$$
2. Strict maximality at the MRC:
$$D_F(\theta_F) = 1, \qquad 0 < D_F(v) < 1 \quad \text{for } v \neq \theta_F.$$
3. Monotonicity from the deepest point: for every $v \in \mathbb{R}$ and $0 \leq \lambda \leq 1$,
$$D_F(v) \leq D_F\big(\theta_F + \lambda(v - \theta_F)\big).$$
4. Vanishing at infinity:
$$D_F(v) \to 0 \qquad \text{as } |v| \to \infty.$$

These conclusions follow from Propositions 1 and 2. In particular, uniqueness gives

$$\theta_{F_Y} = a + b\theta_F,$$

Both $g_F$ and $L_F$ are multiplied by $|b|$, so their ratio is unchanged. If $F$ is symmetric about $c$, reflection symmetry and uniqueness imply $\theta_F = c$, therefore, the symmetry centre is deepest. For each $0 < \alpha \leq 1$, define the central region

$$C_F(\alpha) = \{v \in \mathbb{R}: D_F(v) \geq \alpha\} = \{v \in \mathbb{R}: L_F(v) \leq g_F/\alpha\},$$

These regions are nonempty compact intervals containing $\theta_F$. They are nested as $\alpha$ increases, and $C_F(1) = \{\theta_F\}$. The theorem does not establish a uniquely centred construction for distributions with multiple population MRCs. Such distributions require a separate treatment of centre selection.

## 2.4 Local geometry and boundary imbalance

The original radius function retains information about the local half-mass boundary. Under suitable smoothness conditions, its derivative is determined by the densities at the two endpoints of the corresponding interval.

**Theorem 2.** Fix $v \in \mathbb{R}$, and let $r = G_F(v)$. Suppose that the median of $|X - v|$ is unique, that $F$ has a density $f$ continuous in neighbourhoods of $v - r$ and $v + r$, and that $f(v - r) + f(v + r) > 0$.
Then $G_F$ is continuously differentiable in a neighbourhood of $v$, with

$$G_F{}'(v) = \frac{f(v - r) - f(v + r)}{f(v - r) + f(v + r)}.$$

This follows by applying the implicit function theorem to the half-mass identity. The positive boundary density sum ensures a locally unique radius solution.
A positive derivative indicates greater boundary density at the left endpoint, while a negative derivative indicates greater boundary density at the right endpoint. In particular,

$$-1 \leq G_F'(v) \leq 1$$

If Theorem 2 applies at the MRC, the first order condition gives

$$G_F'(\theta_F) = 0, \qquad f(\theta_F - g_F) = f(\theta_F + g_F).$$

This equality expresses local boundary balance and does not imply symmetry of the full distribution. A minimizing centre may also occur at a nondifferentiable point, where this characterization does not apply. The boundary identity concerns $G_F$, rather than its envelope. Wherever $L_F$ is differentiable,

$$D'_F(v) = -\frac{g_F L'_F(v)}{L_F(v)^2}.$$

Define the envelope adjustment by

$$\Delta_F(v) = L_F(v) - G_F(v) \geq 0.$$

A positive adjustment means that the current radius is smaller than a radius previously encountered along the segment from the MRC. It identifies a departure from outward monotonicity; it does not, by itself, identify the number of modes or clusters.

## 2.5 Empirical construction

For independent observations $X_1, \dots, X_n$, define

$$G_n(v) = \inf\left\{r \geq 0 : \frac{1}{n}\sum_{i=1}^{n} \mathbf{1}\{|X_i - v| \leq r\} \geq \frac{1}{2}\right\}.$$

Thus, $G_n(v)$ is the $k$th smallest absolute distance from $v$, where $k = \lceil n/2 \rceil$. This convention differs from averaging the two middle distances when $n$ is even.
The population minimizer is assumed unique, but the empirical minimizer need not be. More generally, allow an approximate empirical minimizer satisfying

$$G_n(\hat{\theta}_n) \leq \inf_{v \in \mathbb{R}} G_n(v) + \varepsilon_n, \qquad \varepsilon_n \downarrow 0.$$

An exact empirical minimizer corresponds to $\varepsilon_n = 0$. Define

$$\hat{g}_n = G_n(\hat{\theta}_n),$$

$$L_n(v) = \max_{0\le t\le 1} G_n\left(\hat{\theta}_n + t(v - \hat{\theta}_n)\right),$$

and

$$D_n(v) = \frac{\hat{g}_n}{L_n(v)}, \qquad \hat{g}_n > 0.$$

Because the defining segment contains $\hat{\theta}_n$, the empirical score lies in $(0,1]$, equals one at the selected centre and is nonincreasing along each outward ray. Maximality is not guaranteed when the sample minimum is nonunique or the selected centre is approximate.

In one dimension, an exact empirical minimizer can be obtained from the ordered observations. A shortest interval containing $k$ observations are identified by minimizing

$$X_{(i+k-1)} - X_{(i)}, \qquad 1 \le i \le n-k+1.$$

Let $i^*$ be a selected minimizing index. The midpoint of the corresponding interval is an empirical MRC, with

$$\hat{\theta}_n = \frac{X_{(i^*)} + X_{(i^*+k-1)}}{2}, \qquad \hat{g}_n = \frac{X_{(i^*+k-1)} - X_{(i^*)}}{2}.$$

Subject to the stated sample quantile convention, this shortest half construction is consistent with the interval based interpretation of least median location methods(Oja, 2010; Rousseeuw & Leroy, 1987). If several intervals have the same minimum length, a deterministic selection rule should be reported.

Figure 1 compares the empirical median radius function and its centre-outward monotone envelope for samples of size $n = 200$ generated from exponential, multimodal and normal distributions. The right column displays the original normalized reciprocal profile, $\hat{g}_n/G_n(v)$, together with the envelope-based score $D_n(v)$. Vertical dashed lines identify the selected empirical minimizing centres.

For the exponential sample, the radius profile is asymmetric and nearly coincides with its envelope. In the multimodal example, the original profile exhibits a secondary decrease away from the selected centre. The envelope fills this valley and replaces the corresponding secondary reciprocal peak with a plateau. For the normal sample, the profiles are approximately symmetric, and the envelope introduces relatively small adjustments.

These examples distinguish concentration geometry from centre-outward ordering. while the envelope removes outward decreases relative to the selected centre, the original radius profile retains secondary concentration features.

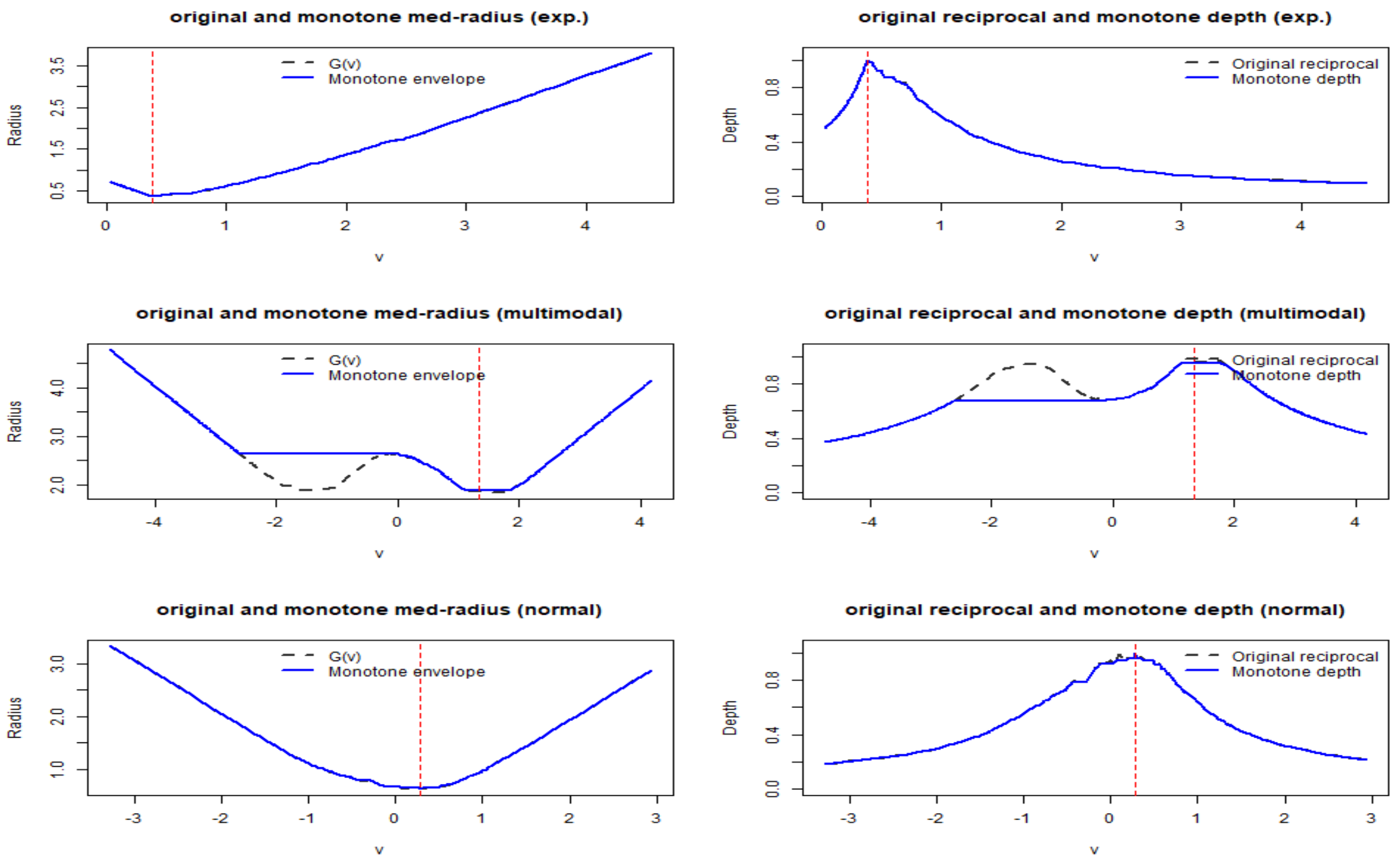

Figure 1. empirical median radius functions and centre-outward monotone envelopes (left column), with their normalized reciprocal profiles (right column) for exponential, multimodal and normal samples of size $n = 200$.

# 3 Multivariate moving median absolute deviation

Let $X \sim F$ be a random vector in $\mathbb{R}^d$ where $F$ is absolutely continuous with respect to Lebesgue measure. Unless otherwise stated, distances are measured using the Euclidean norm.
For each candidate location $v \in \mathbb{R}^d$, define the multivariate moving median absolute deviation by

$$\Phi_F(v) = \text{Med}\|X - v\| = \inf\{r \geq 0: P(\|X - v\| \leq r) \geq 0.5\}.$$

Thus, $\Phi_F(v)$ is the smallest radius of a ball centred at $v$ containing at least half the probability mass. Absolute continuity ensures that

$$P(\|X - v\| \leq \Phi_F(v)) = 0.5.$$

Small values of $\Phi_F(v)$ identify locations around which half the distribution can be enclosed within relatively small balls. This distance quantile criterion is related to least median methods and their extensions to multivariate location (Rousseeuw, 1984).
As in one dimension, the original radius function need not increase monotonically away from its minimizing centre. We therefore retain $\Phi_F$ to describe concentration geometry and construct a radial monotone envelope to obtain a centre-outward ordering.

## 3.1 Basic properties and the median radius center

The following properties extend the univariate results without requiring finite moments.
Proposition 3. The multivariate median-radius function is positive, 1-Lipschitz, and coercive. Specifically,

$$\Phi_F(v) > 0, \qquad v \in \mathbb{R}^d,$$

$$|\Phi_F(v) - \Phi_F(w)| \le \|v - w\|, \qquad v, w \in \mathbb{R}^d,$$

and

$$\Phi_F(v) \longrightarrow \infty \qquad \text{as } \|v\| \longrightarrow \infty.$$

Moreover, if

$$Y = a + bQX,$$

where $a \in \mathbb{R}^d$, $b > 0$, and $Q$ is orthogonal, then

$$\Phi_{F_Y}(a + bQv) = b\,\Phi_F(v).$$

Consequently, $\Phi_F$ attains a strictly positive minimum and its set of minimizing locations is nonempty and compact.

Positivity follows because absolute continuity excludes point masses. The Lipschitz bound follows from the triangle inequality and the ordering of distance quantiles. Coercivity follows because any ball of fixed radius, translated towards infinity, contains probability tending to zero. The transformation property follows from

$$\|Y - (a + bQv)\| = b\,\|X - v\|.$$

**Assumption 2.** The multivariate median radius functional has a unique global minimizer,

$$\theta_F = \arg\min_{v \in \mathbb{R}^d} \Phi_F(v).$$

and

$$\phi_F = \Phi_F(\theta_F) = \min_{v \in \mathbb{R}^d} \Phi_F(v) > 0.$$

The point $\theta_F$ is the multivariate median radius centre (MRC). It is the centre of a smallest Euclidean ball containing at least half the distribution.

The MRC differs from the geometric median that minimizes expected Euclidean distance when that expectation is finite. The MRC minimizes a distance quantile while the geometric median minimizes average distance. Their locations can therefore differ under asymmetric distributions or distributions with separated concentrations.

## 3.2 Radial monotone envelopes

Under Assumption 2, define the radial monotone envelope of $\Phi_F$ by

$$\Psi_F(v) = \max_{0 \le t \le 1} \Phi_F\big(\theta_F + t(v - \theta_F)\big).$$

The closed segment includes both the selected centre and the candidate location. The maximum exists because $\Phi_F$ is continuous.

For a unit direction $u \in S^{d-1}$ and $r \ge 0$, this becomes

$$\Psi_F(\theta_F + ru) = \max_{0 \le s \le r} \Phi_F(\theta_F + su).$$

Thus, the envelope is a running maximum of the original radius along each outward ray.

**Proposition 4.** Under Assumption 2,

$$\Psi_F(v) \ge \Phi_F(v),$$

$$\Psi_F(\theta_F) = \phi_F, \quad \Psi_F(v) > \phi_F, \quad for\ v \neq \theta_F,$$

and

$$\Psi_F(\theta_F + r_1 u) \leq \Psi_F(\theta_F + r_2 u), \qquad 0 \leq r_1 \leq r_2.$$

Furthermore,

$$|\Psi_F(v) - \Psi_F(w)| \leq \|v - w\|,$$

and

$$\Psi_F(v) \longrightarrow \infty \qquad \text{as } \|v\| \longrightarrow \infty.$$

The envelope is the smallest majorant of $\Phi_F$ that is nondecreasing along rays from $\theta_F$. If the original radius is already nondecreasing along every outward ray, we have

$$\Psi_F(v) = \Phi_F(v) \qquad \text{for all } v \in \mathbb{R}^d.$$

The adjustment

$$\Delta_F(v) = \Psi_F(v) - \Phi_F(v) \geq 0,$$

quantifies the increase required to impose radial monotonicity. Positive values indicate secondary decreases in the original radius along a ray. They do not directly identify the number of modes or clusters.

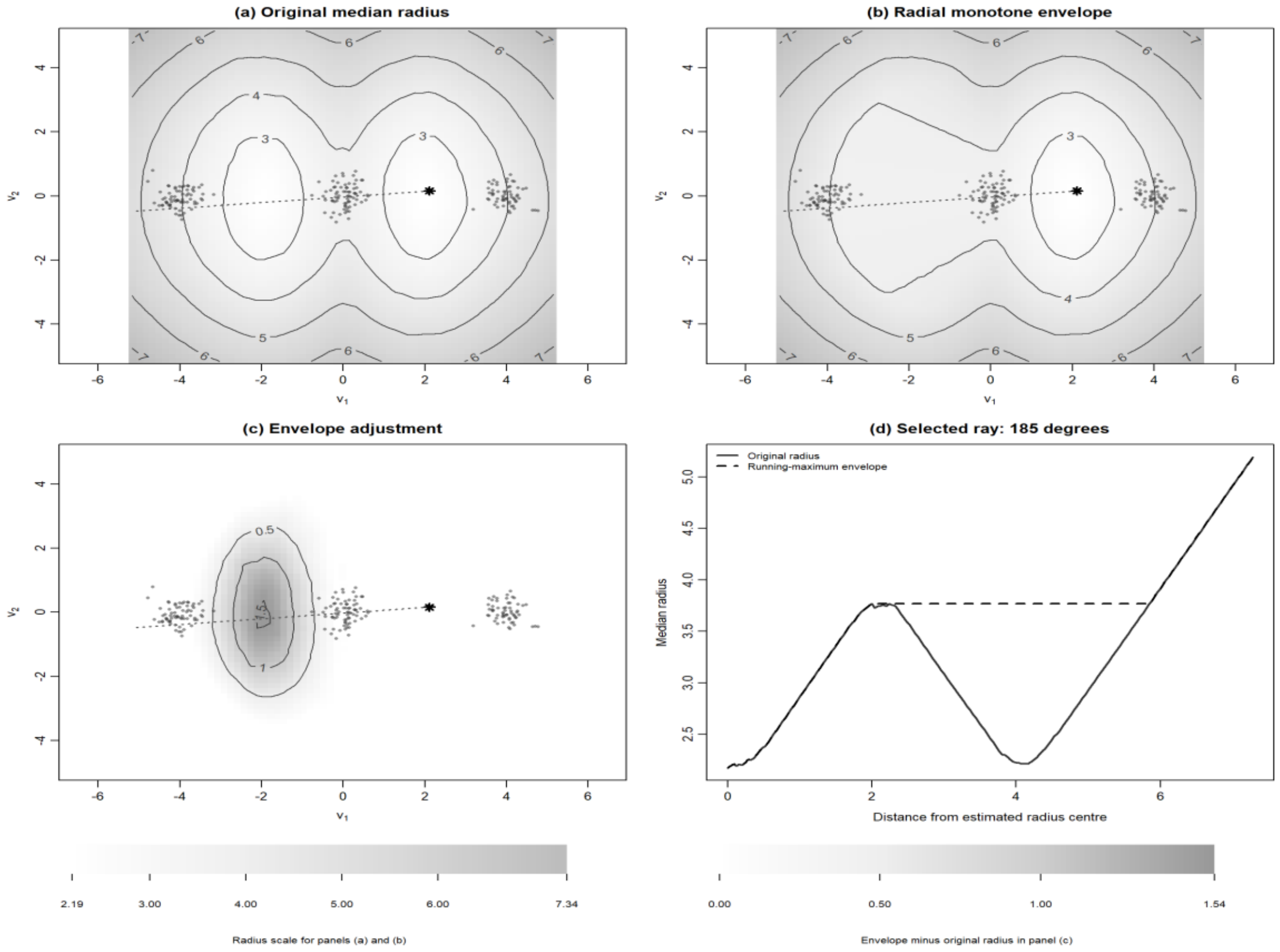

Figure 2. radial monotone envelope construction for a three component mixture sample of size $n = 200$.

Figure 2 illustrates the distinction between the original concentration geometry and its centre-outward modification. The original radius exhibits separated low radius regions that need not coincide with individual mixture component centres. The envelope modifies a secondary valley encountered along rays from the selected empirical MRC. Along the selected direction of 185°, the envelope retains the largest preceding radius while the original radius decreases that producing a plateau in the envelope. The two profiles coincide again once the original radius exceeds that retained value.

## 3.3 Multivariate envelope depth

Define the multivariate envelope-based MMAD depth by

$$D_F(v) = \frac{\phi_F}{\Psi_F(v)}.$$

The following result establishes centre-outward properties analogous to those considered in the general depth framework of (Serfling & Zuo, 2000). For Euclidean distance, the transformation property is similarity invariance.

**Theorem 3.** Under Assumption 2, $D_F$ is continuous and satisfies:

Maximality at the MRC:

$$D_F(\theta_F) = 1, \qquad 0 < D_F(v) < 1 \quad \text{for } v \neq \theta_F,$$

Monotonicity,

$$D_F(v) \leq D_F\big(\theta_F + \lambda(v - \theta_F)\big), \qquad 0 \leq \lambda \leq 1,$$

Vanishing at infinity:

$$D_F(v) \longrightarrow 0 \qquad \text{as } \|v\| \longrightarrow \infty.$$

Similarity invariance: for $Y = a + bQX$,

$$D_{F_Y}(a + bQv) = D_F(v),$$

These properties follow from Propositions 3 and 4. In particular, the minimizing centre transforms as $a + bQ\theta_F$ while the minimum radius and its envelope are both multiplied by $b$. If $F$ is centrally symmetric about $c$, the uniqueness implies $\theta_F = c$ and hence $D_F(c) = 1$.
For $0 < \alpha \leq 1$, define the central region

$$R_F(\alpha) = \{v: D_F(v) \geq \alpha\} = \left\{v: \Psi_F(v) \leq \frac{\phi_F}{\alpha}\right\}.$$

These regions are nonempty, compact, nested and star shaped about $\theta_F$. In particular,

$$v \in R_F(\alpha) \;\Rightarrow\; \theta_F + t(v - \theta_F) \in R_F(\alpha), \qquad 0 \leq t \leq 1.$$

Convexity is not guaranteed. Unlike the one-dimensional construction, radial monotonicity in several dimensions does not imply convex central regions.
Therefore, the Euclidean construction provides a continuous and similarity invariant centre-outward depth on the stated distribution class.

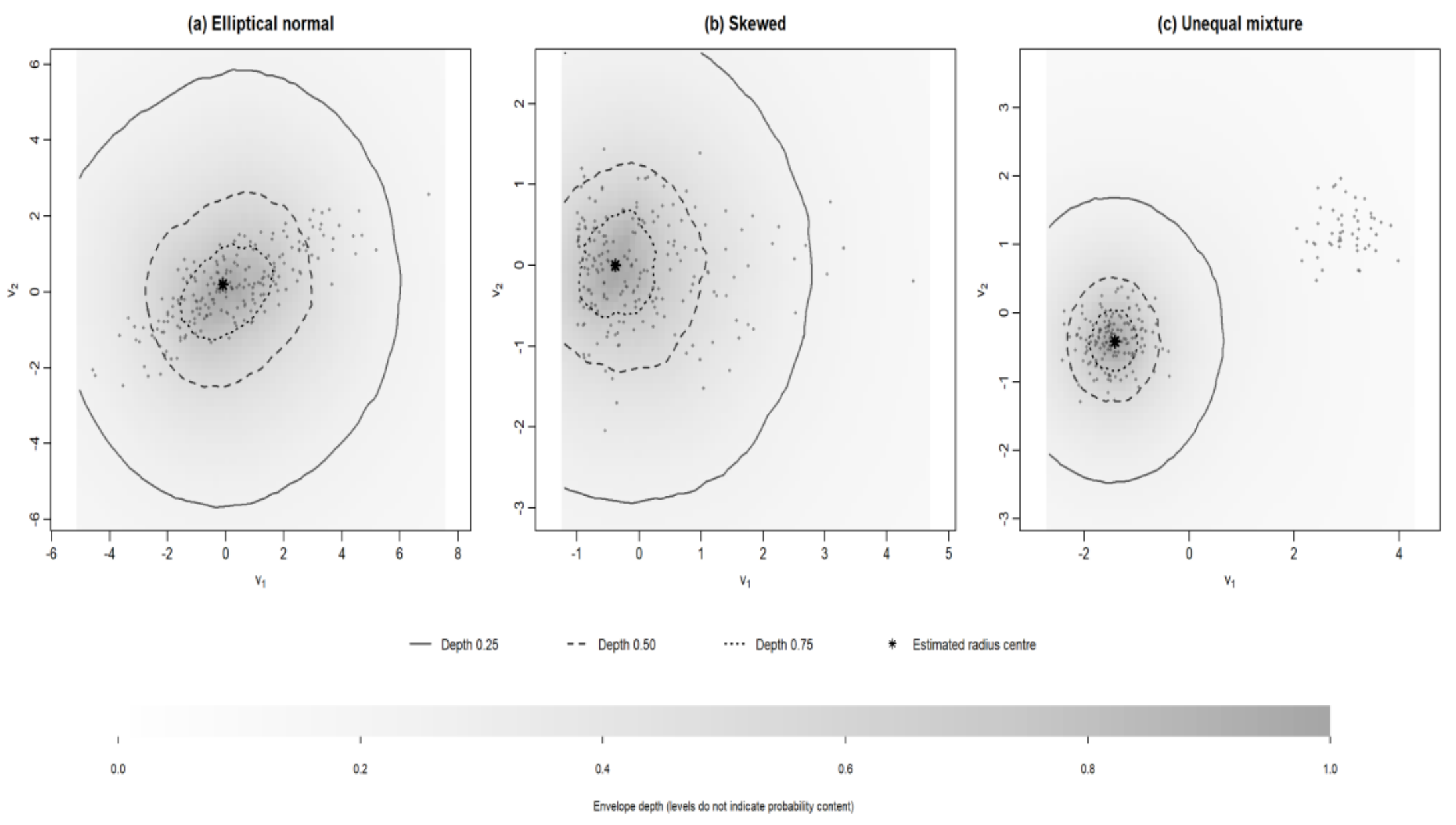

Figure 3. envelope-based depth for elliptical normal, skewed, and unequal mixture samples of size $n = 200$.

Figure 3 presents the envelope-based depth for samples of 200 observations generated from elliptical normal, skewed and unequal mixture distributions. The common grayscale represents depth values with darker shading indicating greater centrality relative to the selected median radius centre.

For the elliptical normal sample, the inner contours are approximately elongated in the direction of the observed dependence while the outer contour is more rounded. This illustrates that Euclidean median radius depth need not produce contours with the same shape at every depth level. For the skewed sample, the contours extend farther to the right of the selected centre that reflecting asymmetric half mass concentration.

For the unequal mixture, the highest depth values occur near the dominant concentration. The smaller, separated component receives lower centrality relative to the selected centre. This behaviour illustrates the construction's emphasis on half mass concentration and should not be interpreted as a classification of the smaller component as contamination.

Across the examples, the displayed contours illustrate nested central regions. Their centre-outward ordering follows from the envelope construction.

## 3.4 Empirical construction

For independent observations $X_1, \ldots, X_n$, define

$$\Phi_n(v) = \inf\left\{r \geq 0 : \frac{1}{n}\sum_{i=1}^{n} \mathbf{1}\{\| X_i - v \| \leq r\} \geq \frac{1}{2}\right\}.$$

This is the $k$th smallest distance from $v$, where $k = \lceil n/2 \rceil$. The same lower median convention should be used throughout the numerical implementation.

Allow an approximate empirical minimizing centre satisfying

$$\Phi_n(\hat{\theta}_n) \leq \inf_{v \in \mathbb{R}^d} \Phi_n(v) + \varepsilon_n, \qquad \varepsilon_n \downarrow 0.$$

Define

$$\hat{\phi}_n = \Phi_n(\hat{\theta}_n),$$

$$\Psi_n(v) = \max_{0\le t\le 1} \Phi_n\left(\hat{\theta}_n + t(v - \hat{\theta}_n)\right),$$

and

$$D_n(v) = \frac{\hat{\phi}_n}{\Psi_n(v)}, \qquad \hat{\phi}_n > 0.$$

The empirical score equals one at the selected centre, lies in (0,1] and is nonincreasing along every outward ray (provided that $\hat{\phi}_n > 0$). Maximality need not hold when the empirical minimum is nonunique or the selected centre is approximate.

Multivariate centre estimation generally requires numerical optimization. A search from several starting points can reduce sensitivity to local minima but does not certify global optimality, uniqueness or the stated approximation bound. These distinctions should be retained when interpreting empirical results. For a scatter-based metric, a positive definite scatter matrix is estimated once and held fixed throughout centre estimation and evaluation of the radius profile. The envelope is approximated by evaluating the radius along a finite grid on each defining segment.

# 4 Conclusion

This study develops a framework based on the moving median absolute deviation for describing half mass concentration and constructing centre-outward orderings. The MRC identifies the location around which half the distribution can be enclosed within the smallest radius. Because the original radius function need not increase monotonically away from this centre, a radial running maximum envelope is introduced to impose outward monotonicity while retaining the original profile for geometric analysis.

Under the stated continuity and unique minimizer assumptions, normalization by the minimum radius produces a continuous score with maximality at the population MRC and vanishing values at infinity. In several dimensions, Euclidean distance yields similarity invariance and the resulting central regions are compact, nested and star shaped although convexity is not guaranteed in the multivariate setting.

The univariate derivative identity provides a complementary interpretation through the relative boundary densities of the half probability interval. It describes local boundary imbalance rather than a global measure of symmetry. The distinction between the original radius and its envelope is therefore substantive: while the latter supports an ordering relative to a selected centre, the former describes concentration geometry.

Several questions remain for further investigation. These include constructions allowing multiple population minimizers, sampling distributions and confidence intervals for estimated centres and computational guarantees for multivariate optimization. Further empirical evaluation is also required to establish when envelope depth offers practical advantages over existing measures. The framework provides a basis for these investigations by separating half-mass concentration, boundary geometry and centre-outward centrality.

**Conflict of interest:** The author does not have any conflict of interest.
**Financial support**: This research did not receive any specific grant from funding agencies in the public, commercial, or not-for-profit sectors.
**AI tools:** ChatGPT and Microsoft Copilot were used to improve language, readability and presentation.